\documentstyle[psfig]{mn}

\newif\ifAMStwofonts

\ifoldfss
  \ifCUPmtlplainloaded \else
    \NewTextAlphabet{textbfit} {cmbxti10} {}
    \NewTextAlphabet{textbfss} {cmssbx10} {}
    \NewMathAlphabet{mathbfit} {cmbxti10} {} 
    \NewMathAlphabet{mathbfss} {cmssbx10} {} 
  \fi
  \ifAMStwofonts
    \ifCUPmtlplainloaded \else
      \NewSymbolFont{upmath} {eurm10}
      \NewSymbolFont{AMSa} {msam10}
      \NewMathSymbol{\upi}     {0}{upmath}{19}
      \NewMathSymbol{\umu}     {0}{upmath}{16}
      \NewMathSymbol{\upartial}{0}{upmath}{40}
      \NewMathSymbol{\leqslant}{3}{AMSa}{36}
      \NewMathSymbol{\geqslant}{3}{AMSa}{3E}

      \let\leq=\leqslant 
       
    \fi
  \fi
\fi 

\ifnfssone
  \newmathalphabet{\mathit}
  \addtoversion{normal}{\mathit}{cmr}{m}{it}
  \addtoversion{bold}{\mathit}{cmr}{bx}{it}
  \newmathalphabet{\mathbfit} 
  \addtoversion{normal}{\mathbfit}{cmr}{bx}{it}
  \addtoversion{bold}{\mathbfit}{cmr}{bx}{it}
  \newmathalphabet{\mathbfss} 
  \addtoversion{normal}{\mathbfss}{cmss}{bx}{n}
  \addtoversion{bold}{\mathbfss}{cmss}{bx}{n}
  \ifAMStwofonts
    \ifCUPmtlplainloaded \else
      \UseAMStwoboldmath
      \makeatletter
      \new@mathgroup\upmath@group
      \define@mathgroup\mv@normal\upmath@group{eur}{m}{n}
      \define@mathgroup\mv@bold\upmath@group{eur}{b}{n}
      \edef\UPM{\hexnumber\upmath@group}
      \new@mathgroup\amsa@group
      \define@mathgroup\mv@normal\amsa@group{msa}{m}{n}
      \define@mathgroup\mv@bold\amsa@group{msa}{m}{n}
      \edef\AMSa{\hexnumber\amsa@group}
      \makeatother
      \mathchardef\upi="0\UPM19
      \mathchardef\umu="0\UPM16
      \mathchardef\upartial="0\UPM40
      \mathchardef\leqslant="3\AMSa36
      \mathchardef\geqslant="3\AMSa3E

      \let\leq=\leqslant 

    \fi
  \fi
\fi 
\ifnfsstwo
  \DeclareMathAlphabet{\mathbfit}{OT1}{cmr}{bx}{it}
  \SetMathAlphabet\mathbfit{bold}{OT1}{cmr}{bx}{it}
  \DeclareMathAlphabet{\mathbfss}{OT1}{cmss}{bx}{n}
  \SetMathAlphabet\mathbfss{bold}{OT1}{cmss}{bx}{n}
  \ifAMStwofonts
    \ifCUPmtlplainloaded \else
      \DeclareSymbolFont{UPM}{U}{eur}{m}{n}
      \SetSymbolFont{UPM}{bold}{U}{eur}{b}{n}
      \DeclareSymbolFont{AMSa}{U}{msa}{m}{n}
      \DeclareMathSymbol{\upi}{0}{UPM}{"19}
      \DeclareMathSymbol{\umu}{0}{UPM}{"16}
      \DeclareMathSymbol{\upartial}{0}{UPM}{"40}
      \DeclareMathSymbol{\leqslant}{3}{AMSa}{"36}
      \DeclareMathSymbol{\geqslant}{3}{AMSa}{"3E}

      \let\leq=\leqslant 

    \fi
  \fi
\fi 

\ifCUPmtlplainloaded \else
  \ifAMStwofonts \else 
    \def\upi{\pi}
    \def\umu{\mu}
    \def\upartial{\partial}
\fi
\fi

\def\lsun{{L_\odot}}
\def\msun{{M_\odot}}

\title[Variable stars in M3]
{Search for variable stars in the globular cluster M3 }

\author[Kaluzny et al.]
  {J.~Kaluzny$^1$\thanks{E-mail: jka@sirius.astrouw.edu.pl(JK),
rwh@st-andrews.ac.uk(RWH),
cclement@astro.utoronto.ca(CC),
rucinski@cfht. hawaii.edu(SMR)}
  R.~W.~Hilditch,$^{2\star}$ C.~Clement,$^{3\star}$ and
  \newauthor
  S.~M.~Rucinski$^{4\star}$\\
  $^1$Warsaw University Observatory, Al.~Ujazdowskie~4,
   00-478~Warszawa, Poland \\
  $^2$School of Physics and Astronomy,
  University of St.~Andrews, North Haugh, St.~Andrews, Fife, KY16~9SS, 
Scotland\\
  $^3$Department of Astronomy, University of Toronto,
   Toronto, Ontario M5S~3H8, Canada\\
  $^4$Canada France Hawaii Telescope Corp., P.~O.~Box 1597, Kamuela, HI 96743,
USA}

\date{Accepted. Received 1997 Sep 1}
\pagerange{\pageref{firstpage}--\pageref{lastpage}}
\pubyear{1997}

\begin{document}

\label{firstpage}

\maketitle

\begin{abstract}

We describe here results of a photometric time-sequence
survey of the globular cluster M3 (NGC 5272), in a search for
contact and detached eclipsing binary stars.
We have  discovered only one likely eclipsing binary and one
SX~Phe type star in spite
of monitoring 4077 stars with $V<20.0$ and observing 25 BSS.
The newly identified SX~Phe star, V237, shows a light curve with a variable 
amplitude. 
Variable V238 shows variability either with a period of 0.49 d 
or with a period of 0.25 d. 
On the
cluster colour-magnitude diagram, the variable occupies a position
a few hundredths of magnitude to the blue of the base of the
red giant branch. V238 is a likely descendent of a binary blue straggler. 

As a side result we
obtained high quality data for 42 of the previously known
RR~Lyrae variables, including 33 of Bailey 
type ab, 7 type c and 2 double-mode pulsators. 
We used equations that relate the physical properties of RRc stars 
to their pulsation
periods and Fourier parameters to derive masses, luminosities,
temperatures and helium parameters for five of the RRc stars. 
%
We also tested equations that relate [Fe/H], $M_V$ and
temperature of RRab stars to pulsation period and Fourier parameters. 
We derived [Fe/H]$=-1.42$ in good agreement with spectroscopic determinations.
%
One of the RRd stars (V79) has switched modes. In
previous studies, it was classified as RRab, but our observations show
that it is an RRd star with the first overtone mode dominating.
This indicates blueward evolution on the horizontal branch.

\end{abstract}


\section{Introduction}

The subject of binarity of stars in globular clusters (GC's) has evolved in
recent years from an entire neglect of the problem 
(in part due to the sin of taking absence of evidence for the 
evidence of absence), to a very active field, with several cosmological
relevances. Binary stars apparently
play a very important role in the dynamical evolution of the GC's
(Hut et al. 1992) and their relative frequency may drastically alter
the direction and outcome over long time scales.
 On the other hand, combined analyses of photometric
and radial-velocity data for eclipsing binaries can give distances,
ages and masses of stars through pure geometry, without recourse to
the cluster isochrone fitting, which suffers from uncertainties involved 
in the transformation of theoretical models into observational quantities.
We note that although masses of RR~Lyrae-type stars can be estimated 
using the pulsation theory, their usefulness is limited because of 
the poorly understood processes of mass loss occurring during 
evolution on the giant branch.

The photometric survey for eclipsing binaries in M3 which is described
in this paper was intended as a first step in a programme aimed 
at a direct, high accuracy determination of the distance and of the
turnoff mass for this cluster. The second future step would 
involve determination of precise radial velocity curves 
for identified detached binaries.
For hypothetical SB2 binaries located at the top of the M3 main-sequence 
($V\approx 18.5$), it is currently feasible to determine masses of 
components with an accuracy not worse than a few percent.
We expected numerous, previously undiscovered, eclipsing
binaries in M3. Our conviction was based on the fact that 
this cluster harbours a rich population of Blue Straggler stars, whose
unusual properties seem to be related to binarity (see below).
M3 was the first GC in which Blue Straggler
stars (BSS) were identified by Sandage (1953).
52 BSS were discovered in the outer parts of the cluster 
using photographic photometry (Sarajedini \& da Costa 1991;
Fusi Pecci et al. 1992, Buonanno et al. 1994), 
while the total number in M3 may be close to 200,
as estimated on the basis of the HST (WFPC-1) results 
for the central one arc-min field by Guhathakurta et al. 1994.
Recently, Nemec \& Park (1996) reported discovery of 3 SX~Phe stars
and 1 eclipsing binary in their survey of the outer-regions of M3. 

For many years the main interest in the BSS centered on their unusual
properties on the extension of the main sequence, but the high incidence
of variability among them, as pulsating SX~Phe stars and contact
binaries, have further increased their importance. There have been
several excellent reviews of the BSS (Stryker 1993) and of the BSS in
globular clusters in particular (Bailyn 1995), which contain
references to the rich literature in the field. Two mechanisms of
formation of the BSS are now generally accepted: the angular-momentum loss
(AML) in close binaries, which is basically independent of 
the cluster dynamics, but is related to the advanced age of the GC, 
and the collisional mechanism in the cluster cores, 
perhaps involving predominantly wide binaries (Leonard 1989, Leonard
\& Fahlman 1991). The unique case of M3, which is the object of 
our study, gives strong support that two mechanisms are
contributing to the BS formation. Ferraro et al (1993)
discovered 70 new BSS and concluded that M3 is the only cluster with
a non-monotonic radial distribution of the BSS, with two maxima, and a
prominent deficiency between 4 and 8 cluster core radii, i.e. at the
angular distances of $100'' - 200''$ from the center.

The telescope that we used in this programme, the Jacobus Kapteyn
Telescope (JKT), had a relatively small focal scale so that
we were forced to conduct our search for variability 
away from the core, at the angular distances of $280''$ N 
and $290''$ S of the cluster center. 
Thus, we would be able to discover variability of the BSS, 
and of related close binaries, in the outer zone where -- supposedly
-- mostly the BSS formed by the AML exists.
We were able to monitor 25 BSS,
but only one turned out to be a variable star of SX~Phe type. 
Moreover we identified one contact binary being a yellow straggler.

Recent studies have shown that the physical properties
of RR~Lyrae variables can be derived from high precision light curves
using the technique of Fourier decomposition. 
Since our survey has provided accurate light curves for 42 
RR~Lyrae stars in M3, the analysis of these curves will be valuable.

\section{Observations and data reduction}

	The photometry of M3, reported in this paper, was obtained by RWH 
during the interval 1996 March 19 - April 2 with the 
1.0-m. Jacobus Kapteyn Telescope
(JKT) at the Observatorio del Roque de Los Muchachos, La Palma. The camera
was mounted at the f/15 Cassegrain focus, with a TEK $1224\times1224$ CCD
providing a field of 6.7 arcmin square, scale 13.8 arcsec\,mm$^{-1}$, and a
pixel size of 0.33 arcsec on the sky. Each night of observations included
bias frames, flat fields on the twilight sky, observations of BVRI standard
star fields (Landolt 1992), and monitoring, in the V filter, of two fields in
the outer parts of M3. A total of 7 good photometric nights were secured,
with additional data on two partial nights.

	The two fields of M3 were selected to be centred on 280 arcsec N 
and 290 arcsec S of the cluster core. The stellar images were well resolved
across most of each field, with only a $\approx 60$ arcsec zone showing 
the crowded
core region across one edge of each frame. The two fields were observed
alternately, and the autoguider system was employed to ensure that every frame
of each field was centred on the same position. On nights of high photometric
quality, a few frames of each field were secured through a B filter
at air masses less than 1.1, in order to provide a good colour-magnitude
diagram for these regions of the cluster. The V filter monitoring usually
extended for 8 continuous hours in a complete night, from air mass 2.0 through
the local zenith to air mass 1.3. On most nights, the seeing was measured
to be in the range 1.0 - 1.5 arcsec. Integration times of 400 sec, typically,
were used to ensure good quality photometry of the main-sequence 
turn-off region at
$19<V<19.5$, as well as ensuring that the brighter BS region
$17<V<19$, and the blue horizontal branch at 
$15.5<V<16.0$, were not saturated. The readout time for each frame
(3 min.) was severely lengthened by the old computer system still employed
for recording data in 1996. A total of 180 V frames of each of the two
fields were secured during the observing session, of which 176 have provided
reliable data.  

The preliminary processing of the raw CCD frames was done with the
standard routines in the IRAF-CCDPROC\footnote{IRAF is distributed by
the National Optical Astronomical Observatories, operated by the
Association of Universities for Research in Astronomy, Inc., under
contract with the National Science Foundation} package.
The stellar profile photometry was extracted using a set of programs and
UNIX scripts developed by JK. The UNIX scripts make use of the
DAOPHOT/ALLSTAR (Stetson 1987) and DOPHOT (Schechter, Mateo \& Saha
1993) programs. More details about the algorithms used may be found 
in Kaluzny et al. (1995).  
The accuracy of derived photometry varied from night to night 
depending on the sky brightness and seeing conditions. For the
prevailing fraction of frames the random errors were smaller than
0.05 mag for $V<20.0$; see Fig. 1.
Transformation to the standard $BV$ system 
was obtained based on observations of the Landolt (1992) equatorial 
standard fields. 
The following relations were adopted:\\
\begin{eqnarray}
 v=c_{1}+V-0.0246\times (B-V)-0.15\times X \\   
b-v=c_{2}+0.9596\times(B-V)-0.12\times X 
\end{eqnarray}
The colour-magnitude diagram for stars from two 
monitored fields is shown in Fig. 2.    
\section{Search for Variables}
The search for variables was conducted by analyzing light curves for 
stars measured on at least 90 frames. There were 3954 and 4043 stars
fulfilling this criterion in the data bases for fields $N$
and $S$, respectively. The total number of stars with $V<20.0$ whose 
light curves were examined for variability was equal to 4077.
To select potential variables we employed three methods, as described in
detail in Kaluzny et al. (1996). The light curves showing possible
periodic signals were selected for further examination. 
44 certain variables were identified this way. Of these 42 are 
RR~Lyr variables, all previously known. The RR~Lyr variables are 
discussed separately in Sec. 4. Of the remaining two variables 
one is an SX~Phe star and another is a likely eclipsing binary. Both of these 
objects are located in the field S. We have designated these two variables
as numbers 237 and 238,
in continuation with the numbering scheme of the catalogue of Sawyer Hogg 
(1973).
Their basic properties and coordinates
are listed in Table 1.  Finding charts for both newly identified
variables are shown in Fig. 3.  V237 and V238 are included in 
the catalogue of M3 stars published by Buonanno et al. (1994).
In that catalogue V237 is listed as BS8484 while V238 is
listed as star \#1101. 

The two observed fields contain a total of 25 stars which can 
be considered blue
stragglers (we adopted the following limits on colour and magnitude  
for M3 blue stragglers: $0<B-V<0.37$,
$16.5<V<18.7$; see Fig. 2).  The light curves of all blue 
stragglers were carefully examined and none of them, 
with the exception of variable V237, showed any evidence for variability
with the full amplitude exceeding 0.05 mag.

\subsection{Variable V237}
V237 is variable blue straggler (see Fig. 2) belonging to  SX~Phe type stars.
Using the AoV statistic (Schwarzenberg-Czerny 1991), we derived for it a
period $P=0.04010$ d. The light curve phased with that period is presented
in Fig. 4. That light curve is unexpectedly noisy considering the small
formal errors of the photometry. Examination of the time-domain light 
curve of V237 reveals
that its amplitude is in fact modulated. This is demonstrated in Fig. 5
in which we show the light curves of V237 observed on the nights of March 25 
and 26, 1996.  
A possible cause of the variable amplitude of V237
could be the presence of more than one periodicity in its light curve. 
The power spectrum obtained with the CLEAN algorithm (Roberts, Lehar \&
Dreher 1987) shows two significant peaks corresponding to closely spaced
periods: $P_{1}=0.0420$~d and $P_{2}=0.0410$~d. These two periods are
too long and too closely spaced to represent any low
sub-harmonics of the fundamental period of the radial pulsations. 
We may offer two possible explanations for the 
observed changes of the amplitude of the light curve of V237:\\
a) One of the periodicities identified in the power spectrum corresponds
to non-radial pulsations.  It is customarily assumed that SX~Phe stars 
pulsate exclusively in radial modes. However, there are no reasons to 
reject the hypothesis that some of them may exhibit variability related 
to non-radial pulsations. Non-radial pulsations were observed for several 
$\delta$~Sct stars which can be considered PopI counterparts of SX~Phe
stars.  \\
b) The variable exhibits the Blazhko effect which is observed for a
significant fraction of RR~Lyrae stars. 
\subsection{Variable V238}
This star has been given the sequential number V238, in continuation
with the numbering scheme of the catalogue of Helen Sawyer Hogg.
The photometric data at the light maximum
are: $V_{max} = 17.25$, $B-V=0.58$.
The light variations of V238 can be phased 
with two possible periods: $P_1 = 0.4983$ day and 
$P_2 = 0.2487$ day. 
The corresponding light curves are shown in
Figure~6. One can see that adoption of the longer period leads 
to a smaller scatter of the phased light curve. 
The colour index at maximum light and the short period 
suggest a contact binary of the W~UMa-type (EW). However, the period of 
0.25 day is very short for the observed colour, if the star 
follows the standard period-colour relation for contact
binaries; the period of 0.49 day seems to be more
likely. The light curve is poorly covered when the longer period 
is assumed; especially poorly covered is the secondary minimum, which
seems to be relatively shallow implying a poor-thermal-contact or
semi-detached binary (EB type). 

V238 is a very interesting and somewhat mysterious binary,
irrespective of whether it is a member of M3 or lies in front of it. 
If it is a member of
the cluster, then its position on the CMD, close to the giant branch,
is unusual. We do not know any EW or EB binaries with comparably 
short orbital periods; such periods would be hard to reconcile with giant
star dimensions. We also do not know any spotted giants of the FK
Comae type which would rotate that rapidly, at the brink of the
rotational breakup. The assumption that the star is a contact binary with
$P = 0.498$ day leads -- via the $M_V (\log P, B-V)$ calibrations 
(Rucinski 1994, 1995) -- to a prediction of $M_V = 3.7 - 3.9$ for 
the cluster parameters of $[Fe/H] = -1.57$ and $E_{B-V} = 0.01$. 
The metallicity correction here is large, about
+0.5 mag; therefore, if the binary is not in the cluster and has solar
metallicity, then $M_V = 3.2 - 3.4$. This is a prediction for a
genuine EW contact system; if this is an EB system, then the star is
probably fainter, as the $M_V$ calibrations tend to over-estimate
luminosities in such situations (Rucinski \& Duerbeck 1997). These
numbers disagree with the absolute magnitude of V238 if it is a
cluster member: With $m-M=14.96$ (Harris 1996), 
it should have $M_V = +2.3$. Thus, it
appears that the binary is projected onto the cluster. This is not
a trivial resolution of the mystery, as at $b = +77^{\circ}$, the implied
distance from the galactic plane would be $z > 6$ kpc. This would make
V238 not only the most distant of known galactic contact binaries, but
also a most distant one from the galactic plane.

\section{RR~Lyrae Variables}
Among the 
42 RR~Lyrae variables that we rediscovered were 33 of Bailey type ab 
(fundamental mode pulsators), 7 type c (first overtone mode) and
two double-mode pulsators  (RRd stars). 
The variables, their Bailey types
and periods  are listed in Table 2; the designation N (north) or S (south)
indicates  the field in which the variable was located.
For the most part, the periods we adopted 
are either those listed by Szeidl (1965) in his major study of
the cluster, or within 0.001 day of those periods. 
(The only exceptions are V203 which was not studied by Szeidl
and V79, the star that has switched modes (cf. section $4.3$.) 
Also listed in the table are the mean
$V$ magnitudes and  the $V$ amplitudes determined by fitting the light curve of
each star to a 6th order Fourier series, i.~e. an equation of the form:  
\begin{eqnarray}
mag=A_0+\sum_{j=1,n}A_j\cos  (j\omega t + \phi_j)
\end{eqnarray}
according to the method of Simon \& Teays (1982). The mean $V$ magnitudes 
listed in Table 2 are
the $A_0$ values that we obtained from the above fits.
For the RRab stars, we also list the maximum deviation parameter $D_m$ as
defined by Jurcsik \& Kovacs (1996, hereafter referred to as JK) for 
establishing whether or not a star has an irregular light curve. The
light curves for the RR Lyrae variables are shown in Figures 7 \& 8. 
Stars with
$D_m>3$ are considered to be irregular.  An examination of the light
curves of Figs.  7 and 8  indicate that, in general this is the case, but there
are some exceptions. In our light curves for V63 and V66, the magnitude of
maximum light is different on different nights, even though the $D_m$ for
both of these stars is
considerably less than 3. Szeidl (1965) also commented that the light
curves for these two stars varied. On the other hand, for V11 and V84, both
with $D_m$ greater than 3, our light curves and Szeidl's study do not reveal
any variations in the light curve.

The period-amplitude relation is plotted in Figure 9 where the symbols 
are as follows: open circles for the RRab stars with 
$D_m>3$, 
solid circles for the those with $D_m<3$, open triangles 
for the RRd stars (plotted with their overtone periods) and 
solid triangles for the RRc stars. There are 
some interesting features to note in the diagram. Among the RRab stars, 
most of the stars with the irregular
light curves have lower amplitudes than the others. These are probably 
`Blazhko variables' where the 
amplitude of the light variation is modulated at a 
longer period than the basic pulsation. 
In  a study of Blazhko variables, Szeidl (1988) found that
the highest amplitude of a Blazhko star fits the period-amplitude relation 
of the regular RRab stars. In the 15 day interval of our observations, we
may have observed some of the Blazhko stars during their phase of maximum
amplitude.  This explains why some of them fit in with the regular RRab
stars in the period-amplitude plot while most of them have lower
amplitudes. One can also see that most of the
regular RRab stars (solid circles) form a tight sequence in the diagram, but
three have high amplitudes for their
periods when compared with the others. These three stars (V14, V65 and V104)
are also the three
brightest of the `regular' RRab stars. This implies that there is a 
period-luminosity-amplitude relation, as previously noted by Sandage (1981a)
in a study of M3. We shall discuss this further in section $4.2$.

In Figure 9, most of the RRc and RRd stars 
form a continuous sequence of decreasing amplitude with increasing
period, but V105 and V203, the two stars with the shortest periods,
do not fit into this sequence. Sandage (1981b) determined colours and
temperatures for some of the RR Lyrae variables in M3 and in his plot of
`reduced' period against temperature, it can be seen that these two stars
have shorter periods than other stars with the same temperature. This
is evidence that the two stars might be second overtone pulsators (RRe stars).
In a recent study of the globular cluster IC 4499, Walker \& Nemec
(1996) found a distribution of amplitude with period similar to Figure 9
and they considered the short period variables to be RRe stars.
However, the models of Bono \it et al. \rm (1997) indicate that there are
short period first overtone pulsators with low amplitudes 
like those of V105 and V203. Thus they could also be RRc stars.
Stellingwerf \it et al.
\rm (1987)  predicted that, if RRe stars exist,  they should have a light 
variation that
has a sharper peak at maximum light than first overtone pulsators. The light
curves in Figure 9 indicate that this may be the case for V203, but not
for V105. 
We have therefore tentatively classified V105 and V203 as RRc stars, but 
note that they may be RRe stars.

\subsection{Fourier Analysis of the RRc Variables}

Using hydrodynamic pulsation models,
Simon \& Clement (1993, hereafter referred to as SC) showed that physical 
properties such as mass, 
luminosity, temperature and a helium parameter
could be computed for RRc stars from the pulsation period and the Fourier
phase parameter $\phi_{31}$ where $\phi_{31}=\phi_3 -3\phi_1$ 
(see equations 2, 3, 6 and 7 of their paper). In an
application of their method to photographic data for RRc stars in six
galactic globular clusters, they found that mean masses and luminosities
increase and mean temperatures fall with decreasing cluster metallicity.
Since the original study of SC, CCD observations of RR Lyrae variables
in a number of clusters have been published
and this has made it possible to calculate Fourier parameters with higher
precision than those derived from photographic data. Subsequent studies
based on CCD data (Clement \it et al. \rm 1995, Clement \& Shelton 1996, 1997)
confirm SC's original result. 
To continue the endeavour to compute physical parameters for RRc stars
in globular clusters, we apply the technique to the RRc stars that
we observed in M3. For this analysis, we
considered only stars for which the error in $\phi_{31}$ is
$\leq 0.2$.
It is not desirable to have errors larger than this
because an error of $0.2$ in $\phi_{31}$ leads to an
uncertainty of $0.03\msun$ in mass, $0.03$ in magnitude and 20K in temperature.
The errors in $\phi_{31}$ for five of the RRc stars that we have 
studied in M3 are all less than $0.2$. For the other two, V105 and V203, 
which may in fact be RRe stars, the errors 
are somewhat larger, mainly because of their low amplitudes.  
The physical parameters (mass, luminosity, temperature and relative
helium abundance) that we have derived 
are listed in Table 3, along with the $\phi_{31}$
values and their standard errors computed according to the method of
Petersen (1986) and the temperatures derived by Sandage (1981b) from the
[$<B>-<V>$] colour.

In Figure 10, we show a plot of $\log L/\lsun$ against mean $V$. The stars
in the north and south fields are plotted with different symbols because
there could be a systematic difference of up to $0.02$ in $V$ magnitude
between the two fields. 
The envelope lines have a slope of $0.4$ and are separated by $ 0.04$
in $\log L$, which represents the uncertainty in the values of $\log L$
computed from $\phi_{31}$ and period. All of the
points fit between the two lines and so our analysis indicates that 
SC's equation gives the correct ranking for $\log L$.

The masses we list for these stars in Table 3
can be compared with masses for the RRd stars in M3.
Nemec \& Clement (1989) analysed published observations of the two
RRd stars V68 and V87. For V68, they  
determined a fundamental period $P_0=0.4785$
and period ratio $P_1/P_0=0.7439$, but for V87, they could not
determine a definitive value for the period ratio because of an
uncertainty in the value of $P_1$.  From the Petersen
diagram of Cox (1995), we estimate a mass of $0.65\msun$ for V68 if
$X=0.7$ and $0.69\msun$ if $X=0.8$. This is comparable to the masses we have
derived from SC's equations for V107 and V75, but somewhat larger than the 
masses for the other stars, particularly V97 and V126 for which we
derived $0.52\msun$.  Evolutionary 
models (Dorman \it et al. \rm 1993; 
Yi \it et al. \rm 1993) do not predict masses as low as $0.52\msun$
in the instability strip. However, de Boer \it et al. \rm (1997) have
estimated a mean mass of $0.38\msun$ for some field HB stars, a value that
is also low when compared with evolutionary models. They derived
these HB masses from $T_{eff}$, $\log g$, luminosity and Hipparcos parallaxes.
Thus the masses that we have derived from SC's equation are comparable to
those determined by other methods.

From Table 3, we see that the temperatures that
Sandage (1981b) derived 
are different from the
ones we computed from $\phi_{31}$, but both methods indicate that V12 is the 
hottest and V126 is the coolest
and the coefficient of correlation between the two sets
of values is $0.75$. 
Sandage also derived temperatures from $<B-V>$. These  
are about 100K lower than the ones he derived from $<B>-<V>$, but they are
highly correlated with them (coefficient of correlation $R=0.999$).
Smith (1995) has pointed out that it is difficult to
establish the effective temperature for a pulsating star to better than
300K. 

In Table 4, we 
compare the mean parameters that we have derived for the M3 RRc stars
with those determined for other clusters. This is an update
to Table 6 of Clement \& Shelton (1997) with the mean absolute $V$ magnitudes
($M_V$) and the Oosterhoff type 
included. These mean magnitudes were calculated from $\log L/\lsun$, assuming a
value of $4.79$ for $M_{bol}$ of the sun (Bishop 1996) and using the
bolometric correction formula adopted by Sandage \& Cacciari (1990).
The [Fe/H] values listed are from Jurcsik (1995) and HB refers to
the parameter (B-R)/(B+V+R) of Lee \it et al. \rm (1994). The mean
parameters for each cluster are based only on stars for which the error
in $\phi_{31}$ is $\leq 0.2$. 
In the 
Clement \& Shelton (1997) paper, the parameters that we quoted
for M3 were based on
$\phi_{31}$ values published by Cacciari \& Bruzzi (1993), but these latter
authors did not quote any errors for their Fourier parameters and so we 
have not 
included their data here. The mean values of $\log L/\lsun$, temperature 
and helium 
parameter 
that we derive from our data are the same as those from Cacciari \& Bruzzi
(1993), but our mean mass is $0.03\msun$ less than theirs. This difference 
could occur either because our sample of only five
RRc stars is not a good representation of the RRc stars
in M3 or because the Cacciari \& Bruzzi sample included a number of stars for
which the uncertainty in $\phi_{31}$ was greater than $0.2$. It is important to
have accurate $\phi_{31}$ values for a larger sample of the RRc stars in M3
because the masses of these stars have important implications for stellar
evolution theory. According to Lee \it et al. \rm (1990), the
RR Lyrae variables in the Oosterhoff type II clusters have evolved away from 
the horizontal branch and therefore should have higher luminosities and
lower masses than those in the type I clusters which are on the ZAHB. 
If the mean mass of the RRc stars in M3 turns out to be greater than the
mean mass for M9 and NGC 2298, this will substantiate the Lee \it et al. \rm
(1990) hypothesis.

Using the data of Table 4, Clement (1996) derived a luminosity-[Fe/H] relation:
\begin{eqnarray}
M_V=0.19[Fe/H]+0.82.
\end{eqnarray}
This is plotted in Figure 11.
The line drawn through the points represents the above equation.
The open circle is the mean $M_V$ ($0.45$) that Mandushev \it et al. \rm
(1996) derived for the HB of the globular cluster M55
by main sequence fitting to nearby subdwarfs with known
trigonometric parallaxes. The agreement is excellent, but
the absolute magnitudes calculated from equation (4) are brighter than those
calibrated by statistical parallax (Layden \it et al. \rm 1996) or by the 
Baade-Wesselink method (Fernley 1994, Clementini \it et al. \rm 1995).
However, there are other investigations that predict brighter values
for $M_V$. For example, a study of the RRd stars in the LMC by  Alcock 
\it et al.  \rm (1997) indicates that the RRd stars have $\log L/\lsun$ ranging 
from $1.69$ to
$1.81$. The stars in question probably have [Fe/H] between $-1.6$ and $-2.4$ 
(Alcock
\it et al. \rm (1996) and so their results are in good agreement with the
data in Table 4. The luminosities derived by Sandage (1993) from pulsation
theory  are also similar to our values. 
His  $M_V$-[Fe/H] relation is: 
\begin{eqnarray}
M_V=0.30[Fe/H]+0.94
\end{eqnarray}
It differs in slope from the
one we list because he uses Zinn's (1985) metallicity scale which does
not have as large a range as that of Jurcsik (1995). Using Zinn's scale,
Clement (1996) derived 
\begin{eqnarray}
M_V=0.27[Fe/H]+0.97
\end{eqnarray}
which is similar to Sandage's (1993) relation.
Even with new data available from the Hipparcos satellite, the controversy
over the luminosity of RR Lyrae stars continues.
RR Lyrae magnitudes determined
by Reid (1997) based on trigonometric parallax measurements of local subdwarfs
are considerably brighter that those derived from statistical parallaxes of
RR Lyrae stars (Fernley \it et al. \rm 1997, Tsujimoto \it et al. \rm 1997).
For the RR Lyrae variables in M5, M15 and M68, Reid (1997) derived
$M_V=0.51$, $0.18$ and $0.20$ respectively, values that are even
brighter than the ones we list for these clusters in Table
4. Taking a different approach, Feast \& Catchpole (1997) applied the
Cepheid P-L relation calibrated with Hipparcos trigonometric 
parallaxes to horizontal
branch stars in the LMC and M31 and estimated that $M_V\sim 0.3$ for RR Lyrae
variables with [Fe/H]$=-1.9$ which is also brighter than the Table 4
values. Thus, at the present time, there is still considerable uncertainty
about the absolute magnitudes of RR Lyrae stars. The various
methods for deriving the absolute magnitudes predict
values that span a range of approximately $0.5$ mag for 
the RR Lyrae variables in metal poor clusters 
like M15 and M68 and $0.3$ mag for more metal rich clusters like M5.
These differences could arise because of invalid assumptions in some of
the methods, or they could occur
because the different methods have been applied to stars with
different properties. It is well known that the luminosity of an RR Lyrae star
depends on its evolutionary state. What if
most field RR Lyrae stars are less evolved
than cluster RR Lyrae stars with the same metallicity? This could
explain why the statistical parallax method, which has been applied
only to field stars, indicates lower luminosities than
main sequence fitting from trigonometric parallaxes of subdwarfs, which has 
been applied only to globular cluster stars. These are some of the issues that
need to be addressed before we can have a full understanding of RR Lyrae
luminosities. 

\subsection{Analysis of the RRab Variables}

In a recent series of papers, JK, Kov\'acs \& Jurcsik (1996, 1997; hereafter 
referred to as KJ96, KJ97) and Jurcsik (1997, hereafter refered to as J97)
have shown that metallicity, absolute magnitudes, intrinsic colours and 
temperatures of RRab stars can be expressed as linear combinations of 
period and the low-order amplitudes and phases of the Fourier decomposition
of their light curves.  Specifically, their formulae are
\begin{eqnarray}
[\rm{Fe/H}]= -5.038-5.394P+1.345\phi_{31}
\end{eqnarray}
\begin{eqnarray}
M_V=1.221-1.396P-0.477A_1+0.103\phi_{31}
\end{eqnarray}
\begin{eqnarray}
V_0-K_0=1.585+1.257P-0.273A_1-0.234\phi_{31}+0.062\phi_{41}
\end{eqnarray}
and
\begin{eqnarray}
\log T_{eff}=3.9291-0.1112(V-K)-0.0032[\rm{Fe/H}]
\end{eqnarray}
The formulae are valid only for RRab stars with `regular' light
curves, i.~e. stars for which the maximum deviation parameter $D_m$ is 
less than 3.
According to Table 2, seventeen
of the RRab stars in our sample have `regular' light curves and
so we can use these stars to test the various formulae of JK, KJ96, KJ97 and
J97.
In Table 5, we list the values we derived for $\phi_{31}$, $\phi_{41}$ and 
$A_1$ from equation (3) and
the $M_V$, temperature and [Fe/H] that we computed from 
equations (7), (8) and (10)
for these 17 RRab stars. (At the bottom of the table, we also include
the data for the two stars V11 and V84 which appear to have regular
light curves, but have $D_m>3$.) Before we applied their equations, 
we added $3.14$ to the $\phi_{31}$ values and subtracted $4.71$ from
the $\phi_{41}$ values that we list in Table 5 because our Fourier
parameters were derived from a cosine series, while theirs were from a sine
series.

The mean value of [Fe/H] derived from $\phi_{31}$ for the (first)
17 RRab stars listed
in Table 5 is $-1.42$ with a standard deviation of $0.10$. This is in
good agreement with [Fe/H]=$-1.47$, the value that Jurcsik (1995) 
adopted for M3, based on spectroscopic observations. Thus, our
data vindicate the formula of JK. 
 
In Figure 12, we plot the $M_V$ values calculated from equation (8)
against mean $V$ for the 17 stars of Table 5. The stars in the north
and south fields are plotted with different symbols because there
could be
a systematic difference of up to $0.02$ in $V$ magnitude between the
two fields. The envelope lines, plotted with a slope of unity, are
separated by $M_V=0.1$ which represents the uncertainty in the calibration
of KJ96.  All of the points fall within these boundaries, thus indicating
that equation (8) reproduces the correct ranking of relative
luminosity for the RRab stars in M3. The actual $M_V$ values are fainter than
the one we list for M3 in Table 4 because the zero point for equation (8)
was calibrated with Baade-Wesselink luminosities published by
Clementini \it et al. \rm (1995). The mean $M_V=0.78$ that
we have calculated for these 17 stars is $0.07$
magnitude brighter than the $M_V=0.85$ that Clement and Shelton (1997)
derived by the same method for three RRab stars in the more 
metal rich globular cluster NGC 6171, thus confirming that RR Lyrae in metal
rich systems are fainter. 

An interesting feature of Figure 12 is the fact that there are three
stars that are brighter than the others. They 
have a mean $V$ magnitude of
$15.5678\pm 0.0353$ compared with $15.6944\pm 0.0275$ for the
other 14. These three stars (V14, V65 and V104) are the three
that do not fit into the
sequence of `regular' RRab stars in the period-amplitude plot of Figure 9.
Sandage (1981a)  also noted a scatter in the period-$B$ amplitude relation for
M3 and attributed it to the fact that, as a consequence of
the period-mean density law for pulsating variable stars,
the more luminous stars are expected to have longer periods.
The period-mean density relation for Oosterhoff type I 
compositions derived by Cox (1995) from models calculated with OPAL opacities
is: 
\begin{eqnarray}
\log P_0 = 11.519+0.829 \log L/\log \lsun -0.647 \log M/ \log \msun - 
3.479 \log T_{eff}(K)
\end{eqnarray}
Sandage corrected for the luminosity effect by using the period-mean density
relation to calculate 
a `reduced' period for each star. The `reduced' period is the period a star
would have if its luminosity were the same as the mean luminosity
of all of the RR Lyrae variables in the cluster.
He found that, in
a plot of amplitude against `reduced' period,
the scatter in the period-amplitude plot was considerably reduced. 
At the time of Sandage's (1981a) study, it was assumed that amplitude was
a unique function of temperature, but this assumption is no longer considered
valid. Since we have derived temperatures for the RRab stars with regular
light curves, we can use these temperatures 
to test this. A plot of amplitude versus temperature is
shown in the upper panel of Figure 13. 
The three brightest stars V14, V65 and V104 are plotted
as open circles and the others are solid circles.
The straight line in the diagram is a least
squares fit to the solid circles. 
The diagram indicates that the three bright stars all have low temperatures
compared with other stars with the same amplitude. 
If equation (10) is valid for
calculating temperature, then there is not a unique
amplitude-temperature relation for the RRab variables in M3.  The 
amplitude-temperature relation is a function of luminosity as well.
The models of  Bono \it et al. \rm (1997) also show that the 
amplitude-temperature relation for RRab stars
is a function of mass and luminosity in the sense that  lower masses and/or
higher luminosities cause the amplitude to be higher at a given temperature.
This is exactly what we see in Figure 13. In the lower panel of
the diagram, we plot mean $<V>$ magnitude against temperature. 
The points delineate a typical horizontal branch evolutionary track (cf.
Yi \it et al. \rm 1993).  Most of the stars appear to be on the
ZAHB, but the three brightest
stars are in a more advanced evolutionary state. 
One might therefore expect these three stars to have increasing periods, 
but Szeidl's (1965) study of period changes does not indicate this to
be the case. His analysis, based on $O-C$ diagrams plotted for
observations made over an interval
of approximately 65  years, showed an increasing period for V14, a 
decreasing period for V104 and a fluctuating period for V65.
However, it is generally accepted (cf. Smith 1995)
that most of the observed period changes of RR Lyrae variables are not caused
by evolution on the horizontal branch. Rathbun \& Smith (1997) pointed out
that the observed rates of period change can be too
large and of the wrong sign because of
a period change `noise' that overlies evolutionary period changes. 
Thus the failure to detect period increases 
for V65 and V104 does not mean they have not yet evolved away from
the ZAHB.

As previously noted, the
temperatures in Table 5 were calculated from equation (10) that relates
temperature to the values of $(V-K)$ and [Fe/H] calculated from 
equations (7) and (9). Thus the temperatures have been derived
from the pulsation period and Fourier parameters.
It would be useful to have an independent
derivation of the temperatures for these stars. Sandage (1981b) derived
temperatures 
for seven of these stars (V1, V18, V34, V51, V65, V74 and V90), and also for
V84, from $<B-V>$ and from [$<B>-<V>$]. In Figure 14, we plot 
the temperatures
calculated from the J97 formula against those that Sandage
derived from [$<B>-<V>$].   
The correlation between the temperatures derived by the two 
methods for these eight stars is not very strong (coefficient of correlation
$R=0.54$), and
it is not clear whether or not V65 fits in with the other points in the
diagram. If we exclude V65, the correlation is stronger ($R=0.70$).
Thus it is not clear from our data whether or not 
J97's formula for calculating the effective temperature for RRab stars
from Fourier parameters is valid. 
Since our conclusion that there is not a unique amplitude-temperature
relation for the RRab stars in M3 is based on the temperatures that we
calculated for the three bright stars (V14, V65 and V104) from Fourier
parameters, this conclusion is called into question.

\subsection{The RRd Variables}

The analysis of RRd stars  is often a useful method for determining stellar
masses, but such analyses require that the periods be known to a precision
of 4 figures. Since  our observations were made over an interval of only
two weeks, we can not determine periods for the two RRd stars
with the necessary precision. 
However, what is interesting about these stars is the fact that the strength
of the first overtone oscillations has increased since they were last
observed. 
In fact, previous studies indicated that V79 was an RRab star and now it
is an RRd star with the first overtone dominating.
This result, which has
been discussed in another paper by us (Clement \it et al. \rm 1997),
indicates blueward evolution on the horizontal branch. 
 
\section{Summary}

From a photometric time-sequence survey of 4077 stars (including 25 BSS)
with $V<20$ in the globular cluster M3, we have been unable to find any
definite eclipsing binary stars. 
We have  discovered only one likely eclipsing binary and one
SX~Phe type star in spite
of monitoring 4077 stars with $V<20.0$ and observing 25 BSS.
The newly identified SX~Phe star, V237, shows a light curve with a variable 
amplitude. It may be the first known SX~Phe star exhibiting the Blazhko
effect. An alternative interpretation is that V237
shows non-radial pulsations.  
Variable V238 shows variability either with the period of 0.49 d 
or with period of 0.25 d. Adoption of the longer period leads to
classification of V238 as a close eclipsing binary. On the
cluster colour-magnitude diagram, the variable occupies a position
a few hundredths of magnitude to the blue of the base of the
red giant branch. V238 is a likely descendent of a binary blue straggler. 

As a side result we
obtained high quality data for 42 of the previously known
RR~Lyrae variables, including 33 of Bailey 
type ab, 7 type c and 2 double-mode pulsators. 
We used equations that relate the physical properties of RRc stars 
to their pulsation
periods and Fourier parameters to derive masses, luminosities,
temperatures and helium parameters for five of the RRc stars. 
The derived values for mass, luminosity  and temperature are in reasonable 
agreement
with values obtained by other methods. A comparison of these new values
with the data for six other galactic globular clusters confirms previous
results that indicated that mean masses and mean luminosities increase and
mean temperatures and helium abundances decrease with decreasing cluster
metal abundance. We also tested equations that relate [Fe/H], $M_V$ and
temperature of RRab stars to pulsation period and Fourier parameters. 
We derived [Fe/H]$=-1.42$ in good agreement with spectroscopic determinations.
We also found that the derived $M_V$ values correlated well with the observed
mean $V$ magnitudes.  If the derived temperatures are correct, then our
observations demonstrate that there is not a unique amplitude-temperature
relation for the RRab stars in M3. The amplitude-temperature relation
depends on luminosity. One of the RRd stars (V79) has switched modes. In
previous studies, it was classified as RRab, but our observations show
that it is an RRd star with the first overtone mode dominating.
This indicates blueward evolution on the horizontal branch.
\section*{acknowledgements}

We thank the PATT for the award of telescope time at the Isaac
Newton Group of telescopes at the Roque de Los Muchachos Observatory, 
La Palma, Canary Islands, and the technical staff for their assistance.

RWH thanks the Particle Physics and Astronomy Research Council 
for the award of a research grant in support of this work. 
JK was supported by the
Polish Committee of Scientific Research through grant 
2P03D-011-12 and by  NSF grant AST-9528096 to Bohdan Paczy\'nski.
CMC and SMR were supported by operating grants from the Natural Sciences and
Engineering Research Council of Canada.

\newpage
 
\begin{table*}
\begin{minipage}{110mm}
\caption{Photometric data and coordinates for variable stars V237 \& V238}
\begin{tabular}{@{}lccclccl}
Var &  $V_{max}$ & $V_{min}$ & $(B-V)$ & $P$[days] & X & Y & Type \\
V237  & 17.90 & 18.10  & 0.15 & 0.04010 & 19 & -181 &SX Phe \\
V238  & 17.25 & 17.35  & 0.58 & 0.4983  & 55 & -265 &Ecl? \\
\end{tabular}

\medskip
Remarks:
\begin{itemize}
\item [XY:] 
The rectangular coordinates X \& Y are offsets in arcsec 
from the cluster center. The coordinates are tied to the system of 
rectangular coordinates from the Sawyer Hogg (1973) catalogue.
\item[V238:] Period P=0.25 d is also feasible but excludes
classification of variable as eclipsing binary (see Fig. 6).
\end{itemize}
\end{minipage}
\end{table*}
 
\begin{table*}
\begin{minipage}{110mm}
\caption{Elements of the RR Lyrae Variables}
\begin{tabular}{@{}rclccr}
Variable & Bailey & Period & $<V>$ & Amplitude & $D_m$ \\
\& Field & type   & (days) &       & (V)       &       \\
  1 (S)  &    ab  &    0.5206250 &  15.6899 &  1.1461 &   2.1 \\
 10 (N)  &    ab  &    0.5695185 &  15.6803 &  0.9203 &   1.4 \\
 11 (S)  &    ab  &    0.5078918 &  15.6861 &  1.3357 &   3.5 \\
 12 (S)  &    c   &    0.3178890 &  15.6198 &  0.5156 &       \\
 13 (S)  &    ab  &    0.4799    &  15.6803 &  0.7072 &  38.5 \\
 14 (S)  &    ab  &    0.6359019 &  15.5658 &  0.8556 &   2.0 \\
 15 (S)  &    ab  &    0.5300794 &  15.6686 &  1.1034 &   1.2 \\
 17 (S)  &    ab  &    0.5757    &  15.6939 &  0.6978 &   7.6 \\
 18 (S)  &    ab  &    0.5163623 &  15.7534 &  1.1492 &   2.2 \\
 23 (N)  &    ab  &    0.5953756 &  15.6326 &  0.7293 &   6.9 \\
 34 (N)  &    ab  &    0.5591012 &  15.6743 &  0.9635 &   1.9 \\
 35 (S)  &    ab  &    0.5296    &  15.6530 &  1.1796 &   2.3 \\
 50 (S)  &    ab  &    0.5130879 &  15.6861 &  0.5435 &   9.5 \\
 51 (S)  &    ab  &    0.5839818 &  15.6906 &  0.8461 &   1.9 \\
 52 (N)  &    ab  &    0.5162250 &  15.7067 &  1.1988 &   2.7 \\
 53 (N)  &    ab  &    0.5048878 &  15.7245 &  1.2088 &  19.5 \\
 59 (S)  &    ab  &    0.5888053 &  15.6854 &  0.8337 &   1.5 \\
 61 (N)  &    ab  &    0.5209312 &  15.7021 &  1.2253 &   3.4 \\
 62 (N)  &    ab  &    0.6524077 &  15.6313 &  0.4796 &   5.7 \\
 63 (N)  &    ab  &    0.5704164 &  15.6997 &  0.8489 &   2.4 \\
 64 (N)  &    ab  &    0.6054588 &  15.6972 &  0.6883 &   3.7 \\
 65 (N)  &    ab  &    0.6683397 &  15.5335 &  0.9460 &   2.2 \\
 66 (N)  &    ab  &    0.6191    &  15.6645 &  0.6463 &   1.6 \\
 67 (N)  &    ab  &    0.5683609 &  15.7267 &  0.5535 &  16.3 \\
 68 (N)  &    d   &    0.356     &  15.6359 &  0.348  &       \\
 69 (N)  &    ab  &    0.5665878 &  15.7009 &  0.9097 &   1.2 \\
 70 (N)  &    ab  &    0.4865    &  15.3660 &  0.3434 &  25.2 \\
 74 (N)  &    ab  &    0.4921441 &  15.7237 &  1.2526 &   2.0 \\
 75 (N)  &    c   &    0.3140790 &  15.6544 &  0.4809 &       \\
 79 (N)  &    d   &    0.358     &  15.7122 &  0.313  &       \\
 84 (N)  &    ab  &    0.5957289 &  15.6733 &  0.7246 &   4.1 \\
 90 (S)  &    ab  &    0.5170334 &  15.7305 &  1.1658 &   2.9 \\
 92 (S)  &    ab  &    0.5035553 &  15.7330 &  0.9990 &  18.1 \\
 97 (S)  &    c   &    0.3349289  & 15.6886  & 0.4000 &       \\
104 (N)  &    ab  &    0.5699231 &  15.6040 &  1.3070 &   2.2 \\
105 (N)  &    c?  &    0.2877427 &  15.6014 &  0.3264 &       \\
106 (N)  &    ab  &    0.5471593 &  15.7183 &  0.6389 &   9.8 \\
107 (N)  &    c   &    0.3090351 &  15.6580 &  0.5138 &       \\
118 (S)  &    ab  &    0.4993807 &  15.7424 &  0.8754 &   4.8 \\
124 (S)  &    ab  &    0.7524328 &  15.5423 &  0.3371 &  12.0 \\
126 (S)  &    c   &    0.3484043 &  15.6348 &  0.4019 &       \\
203 (S)  &    c?  &    0.28964   &  15.5877 &  0.1758 &       \\
\end{tabular}
\end{minipage}
\end{table*}

\begin{table*}
\begin{minipage}{110mm}
\caption{Parameters for the RRc stars in M3 calculated from SC's equations}
\begin{tabular}{@{}lccccccc}
Star & $\phi_{31}$  & $\sigma$  & Mass   & $\log L/\lsun$  & Temp  & helium &
Temp  \\
     &              & ($\phi_{31}$) &    &        & ($\phi_{31}$) &  parameter
&   \\
V203 & 5.56 &  0.69 &   ---   &  ---   & ---   &  ---   & 7447\\
V105 & 3.12 &  0.24 &   ---   &  ---   & ---   &  ---   & 7516 \\
V107 & 2.66 &  0.05 &  0.682  & 1.726  & 7304  & 0.265  & 7345 \\
V75  & 2.86 &  0.04 &  0.650  & 1.721  & 7306  & 0.268  & 7244 \\
V12  & 3.24 &  0.07 &  0.595  & 1.704  & 7329  & 0.275  & 7396 \\
V97  & 3.89 &  0.12 &  0.513  & 1.687  & 7340  & 0.284  &  \\
V126 & 3.94 &  0.13 &  0.523  & 1.705  & 7298  & 0.279  & 7063 \\
\end{tabular}
\end{minipage}
\end{table*}

\begin{table*}
\begin{minipage}{110mm}
\caption{Mean parameters for cluster RRc stars}
\begin{tabular}{@{}lccrcccccc}
Cluster & Oosterhoff & [Fe/H] & HB & No.of & mean & mean & mean & mean  &  mean 
\\
        & type &     &    & stars & mass & $\log L/\lsun$ &  temp & helium 
& Mv \\
NGC 6171 & I & -0.68 & -0.76 &  6 &   0.53 &  1.65 &  7447 &  0.29 & 0.65 \\
M5       & I & -1.25 &  0.19 &  7 &   0.58 &  1.68 &  7388 &  0.28 & 0.61 \\
M3       & I & -1.47 &  0.08 &  5 &   0.59 &  1.71 &  7315 &  0.27 & 0.55 \\
M9       & II & -1.72 &  0.87 &  1 &   0.60 &  1.72 &  7299 &  0.27 & 0.53 \\
NGC 2298 & II & -1.90 &  0.93 &  2 &   0.59 &  1.75 &  7200 &  0.26 & 0.47 \\
M68      & II & -2.03 &  0.44 & 16 &   0.70 &  1.79 &  7145 &  0.25 & 0.38 \\
M15      & II & -2.28 &  0.72 &  6 &   0.73 &  1.80 &  7136 &  0.25 & 0.37 \\
\end{tabular}
\end{minipage}
\end{table*}

\begin{table*}
\begin{minipage}{110mm}
\caption{Properties of RRab stars in M3 calculated from the JK/KJ equations}
\begin{tabular}{@{}lcccccc}
Star & $\phi_{31}$ & $\phi_{41}$ &A1 &   Mv  &   Temp & [Fe/H] (error) \\
V1  & 1.54 & 5.73 & 0.3970 & 0.79  & 6484 & -1.55 (0.05)  \\
V10 & 1.81 & 6.09 & 0.3154 & 0.79  & 6408 & -1.45 (0.03)  \\
V14 & 2.14 & 6.62 & 0.3035 & 0.74  & 6335 & -1.37 (0.05)  \\
V15 & 1.60 & 5.83 & 0.3709 & 0.80  & 6464 & -1.52 (0.06)  \\
V18 & 1.67 & 5.85 & 0.3791 & 0.82  & 6514 & -1.35 (0.05)  \\
V34 & 1.91 & 6.03 & 0.3606 & 0.79  & 6486 & -1.26 (0.04)  \\
V35 & 1.63 & 5.91 & 0.4271 & 0.77  & 6492 & -1.48 (0.06)  \\
V51 & 1.98 & 6.35 & 0.2856 & 0.80  & 6397 & -1.30 (0.03)  \\
V52 & 1.58 & 5.62 & 0.4415 & 0.78  & 6537 & -1.47 (0.07)  \\
V59 & 1.99 & 6.31 & 0.3015 & 0.79  & 6403 & -1.31 (0.04)  \\
V63 & 1.79 & 6.12 & 0.3195 & 0.79  & 6399 & -1.48 (0.08)  \\
V65 & 2.18 & 6.74 & 0.3243 & 0.69  & 6287 & -1.49 (0.04)  \\
V66 & 2.08 & 6.63 & 0.2390 & 0.79  & 6317 & -1.36 (0.04)  \\
V69 & 1.90 & 6.26 & 0.3041 & 0.81  & 6419 & -1.32 (0.03)  \\
V74 & 1.57 & 5.81 & 0.4278 & 0.82  & 6552 & -1.36 (0.06)  \\
V90 & 1.65 & 5.80 & 0.3847 & 0.81  & 6513 & -1.38 (0.05)  \\
V104& 1.69 & 5.89 & 0.4241 & 0.72  & 6438 & -1.62 (0.06)  \\
\\
V11 & 1.57 & 5.68 & 0.4402 & 0.79  & 6542 & -1.44 (0.05)  \\
V84 & 2.09 & 6.53 & 0.2543 & 0.81  & 6389 & -1.22 (0.03)  \\
\end{tabular}
\end{minipage}
\end{table*}
\clearpage

\setcounter{figure}{0}
\begin{figure}
\hspace{2cm}
\centerline{\psfig{figure=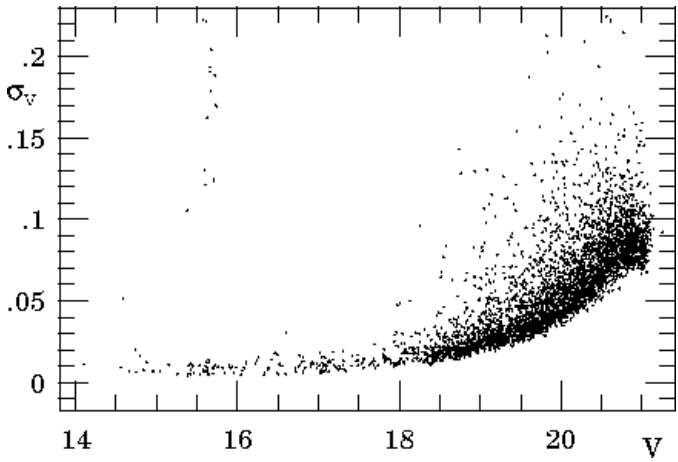,height=6cm}}
\caption{
The single-measurement errors of our photometry for
field N versus the $V$ magnitudes for 110 of our best $V$-filter frames.
The group of stars with large values of rms present at $V\approx 15.7$  
are RR~Lyrae variables.
}
\end{figure}
\setcounter{figure}{1}
\begin{figure}
\vspace{2cm}
\centerline{\psfig{figure=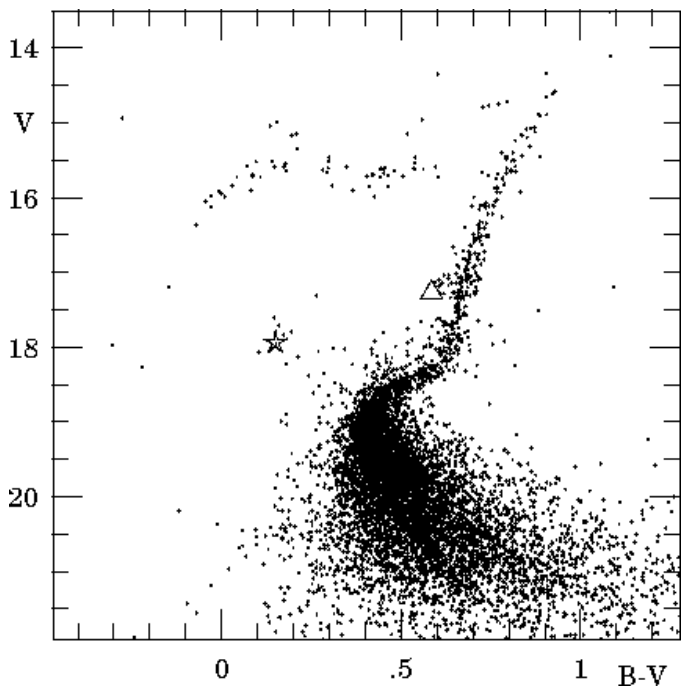,height=9cm}} 
\caption{
The colour-magnitude diagram for stars from two
monitored fields. SX Phe variable V237 is marked with an asterisks while
contact binary V238 is marked with an open triangle. RR~Lyrae variables are
not plotted.
}
\end{figure}
\clearpage
\setcounter{figure}{2}
\begin{figure}
\hspace{3cm}
\psfig{figure=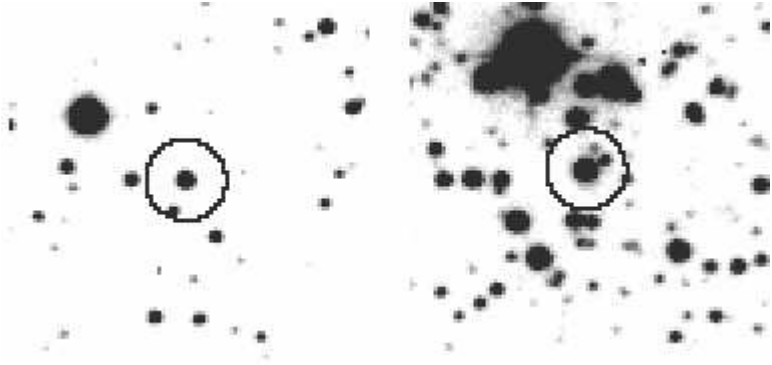,height=5cm}
\caption{
The $V$-band finding charts for variables 
V237 (left) and V238 (right). Each chart is 45 arcsec on a side with
north up and east to the left.
}
\end{figure}
\setcounter{figure}{3}
\begin{figure}
\hspace{4cm}
\psfig{figure=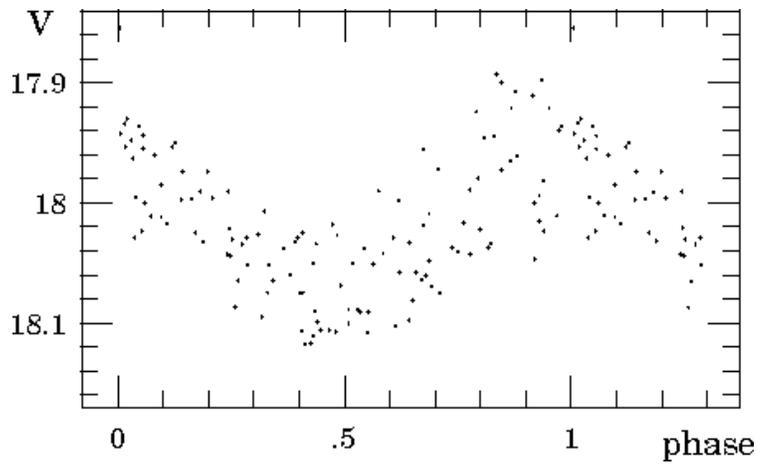,height=6cm}
\caption{
The light curve of SX~Phe variable V237 phased with a
period $P=0.04010$ d.
}
\end{figure}
\clearpage
\setcounter{figure}{4}
\begin{figure}
\hspace{3.5cm}
\psfig{figure=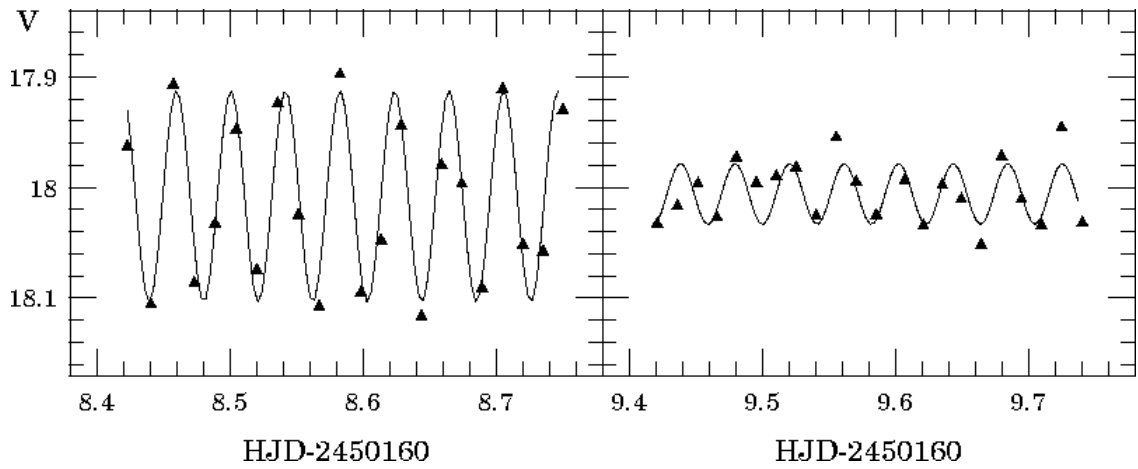,height=6cm}
\caption{
The light curves of variable V237 observed on March 25
(left panel) and March 26, 1996. The sine with a period of 0.04010 d
is superposed on both light curves. 
}
\end{figure}
\setcounter{figure}{5}
\begin{figure}
\vspace{2cm}
\centerline{\psfig{figure=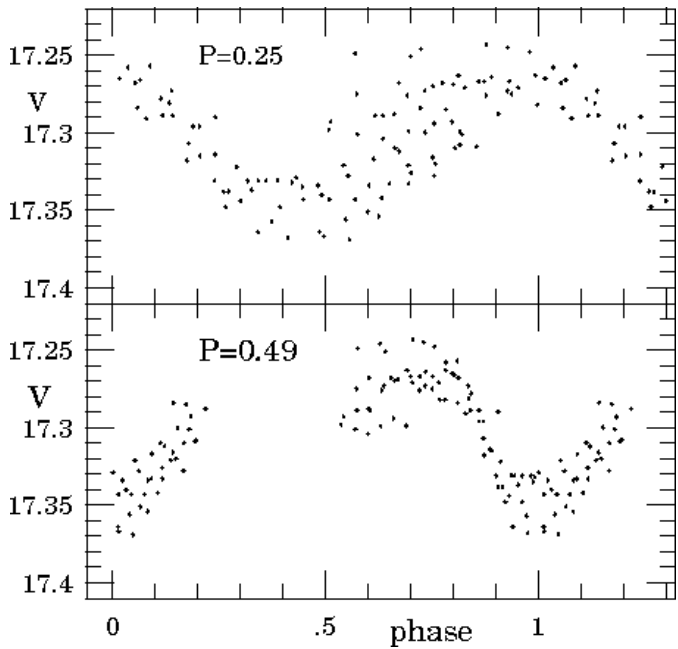,height=8.5cm}} 
\caption{
The light curve of the variable  V238 phased with 
period $P=0.4983$ d (lower panel) and $P=0.2487$ d (upper panel).
}
\end{figure}
\clearpage
\setcounter{figure}{6}
\begin{figure}
\hspace{4cm}
\psfig{figure=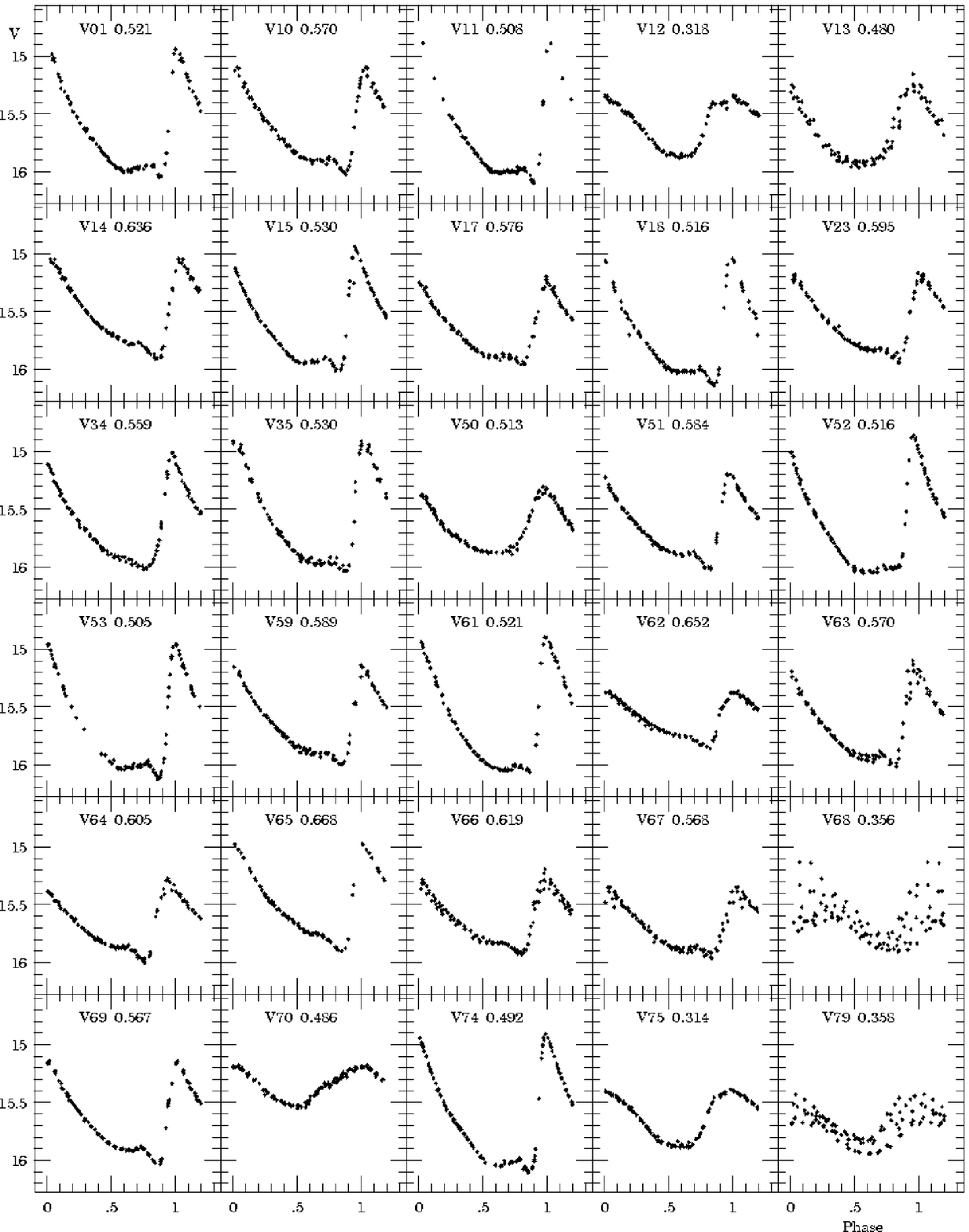,height=22.3cm}
\caption{
The $V$-band light curves for RR~Lyrae  variables 
V1-79 plotted with the periods listed in Table 2.
}
\end{figure}
\clearpage
\setcounter{figure}{7}
\begin{figure}
\hspace {4cm}
\psfig{figure=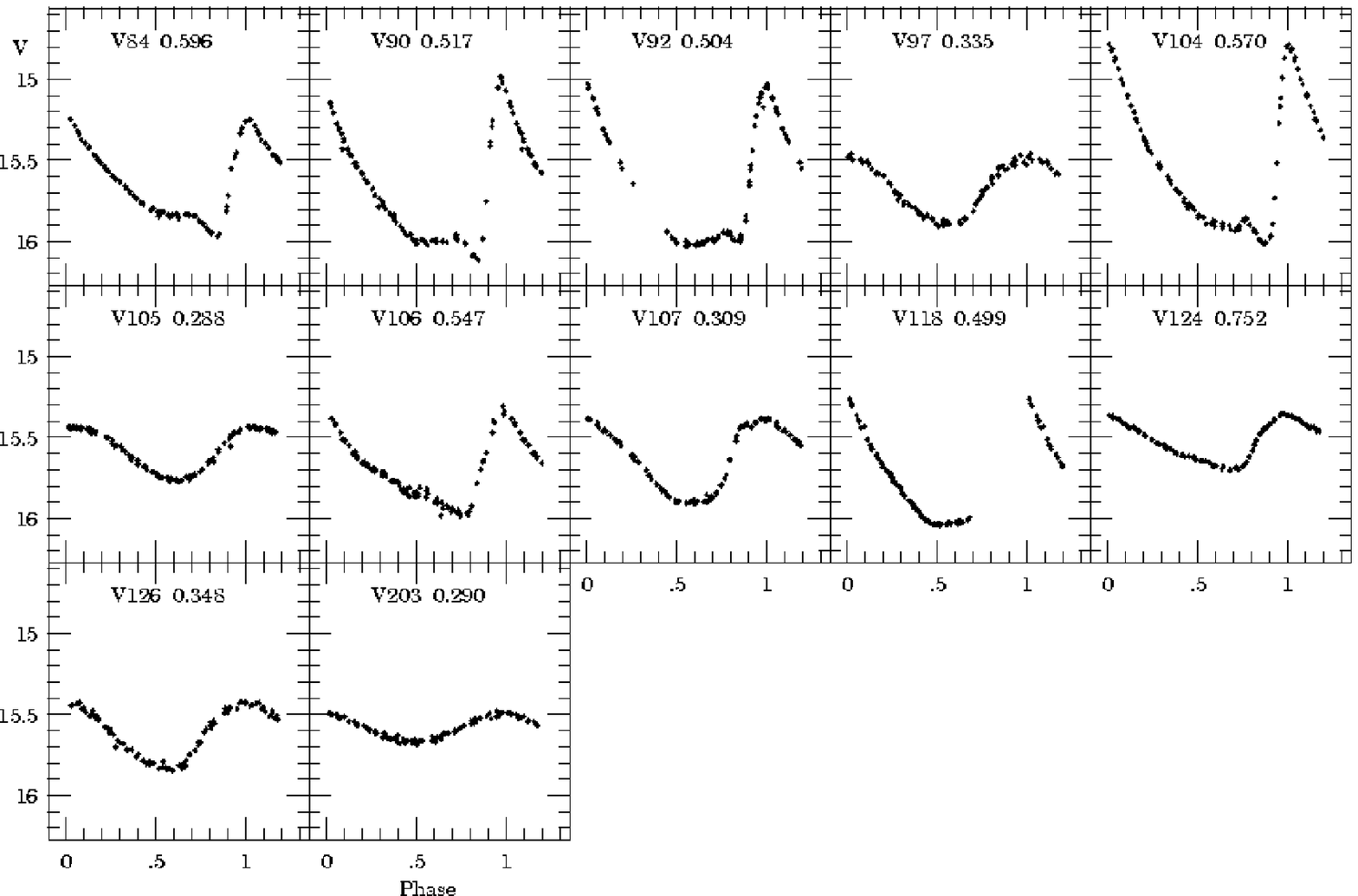,height=12cm}
\caption{
The $V$-band light curves for RR~Lyrae  variables 
V84-203  plotted with the periods listed in Table 2.
}
\end{figure}
\clearpage
\setcounter{figure}{8}
\begin{figure}
\hspace{2cm}
\psfig{figure=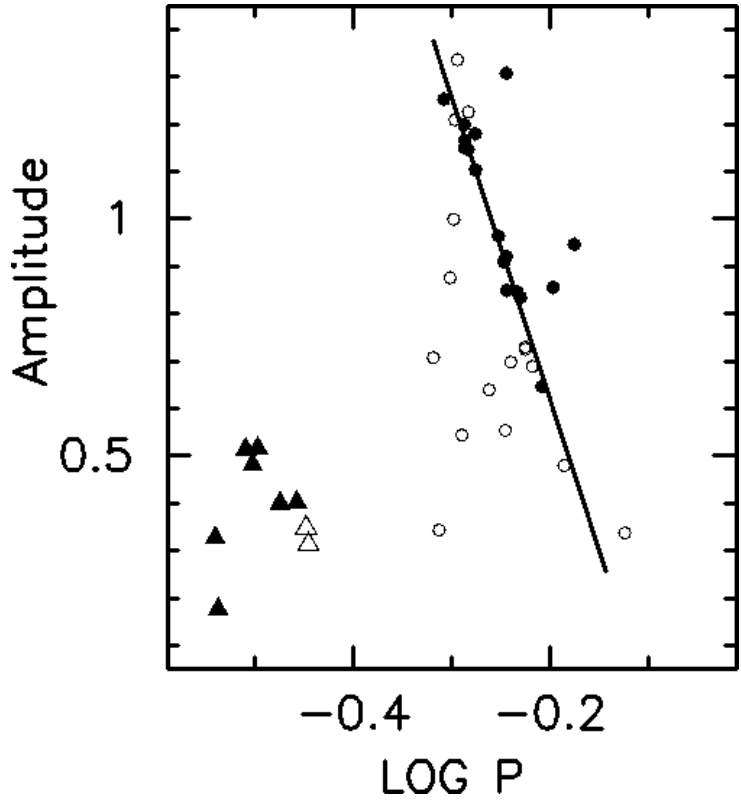,height=10.5cm}
\caption{
The period-amplitude relation ($A_V$ versus $\log P$)
for the 42 RR~Lyrae
variables. The symbols are as follows: solid circles for ordinary RRab stars,
open circles for RRab stars with irregular light curves according to the
$D_m$ parameter,
solid triangles for RRc stars and
open triangles for RRd stars which have been plotted with their first overtone
periods. The straight line was derived from a least 
squares fit to the 14 RRab stars for which $D_m<3$ and mean $V>15.64$.
}
\end{figure}
\clearpage
\setcounter{figure}{9}
\begin{figure}
\hspace{2cm}
\centerline{\psfig{figure=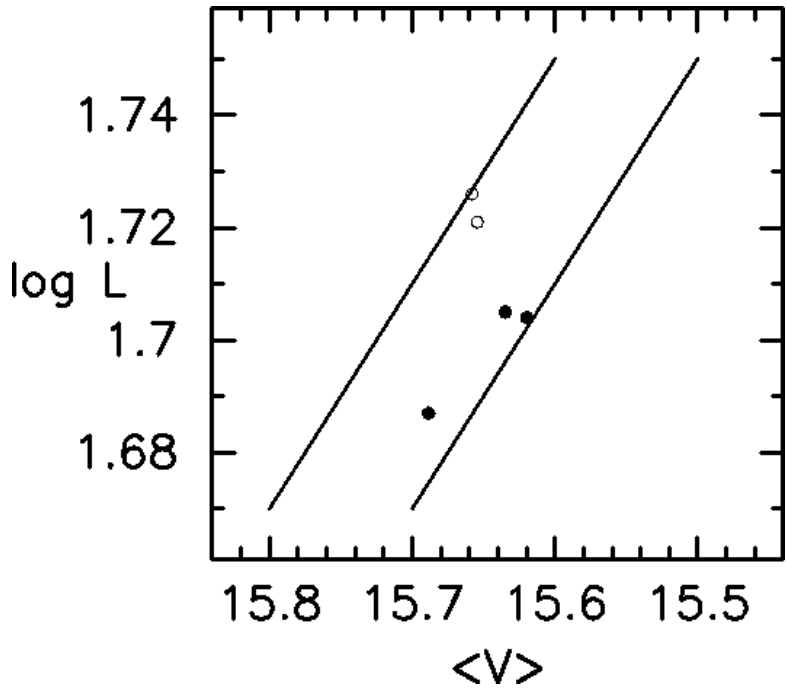,height=9cm}} 
\caption{
$\log L/\lsun$ derived from SC's equation relating
luminosity to period and $\phi_{31}$, plotted against mean $V$
for the RRc stars in M3. The open circles represent the stars 
in the field
north of the cluster centre and the solid circles represent the stars
south of the cluster centre. The envelope lines have a slope of
$0.4$ and are separated by $\log L = 0.04$ which represents the uncertainty
in the fit to the models. 
}
\end{figure}
\clearpage
\setcounter{figure}{10}
\begin{figure}
\hspace{2cm}
\centerline{\psfig{figure=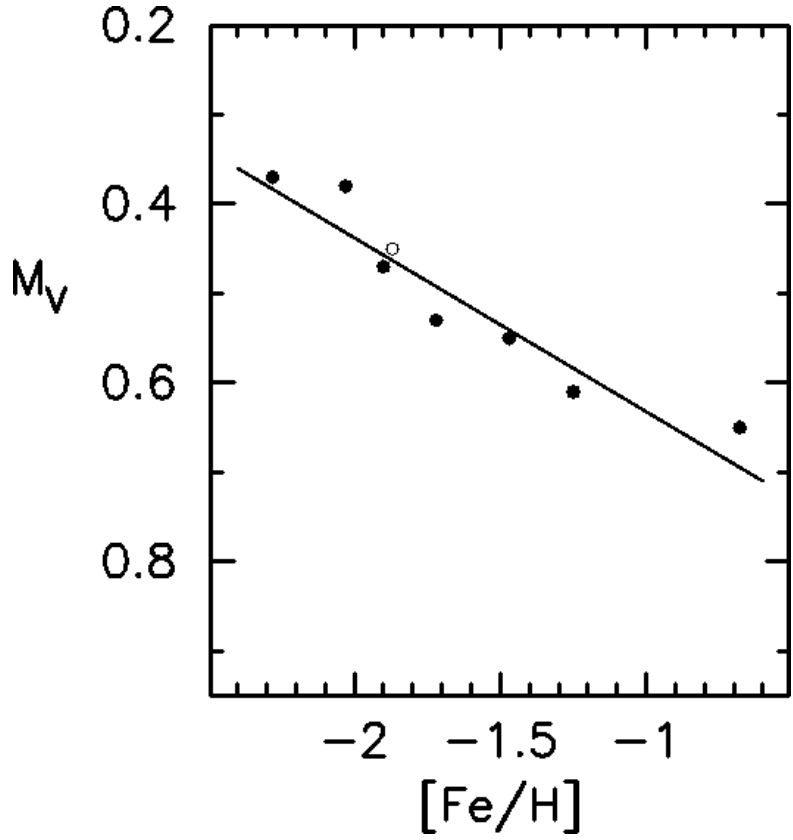,height=11cm}} 
\caption{
A plot of $M_V$ against [Fe/H] using the data from Table
4 (solid circles). The line fit through the points is our relation: 
$M_V=0.19[\rm{Fe/H}] +0.82$.
The open circle represents the mean $M_V$ that Mandushev \it et al. \rm (1996)
derived for M55 by main sequence fitting.
}
\end{figure}
\clearpage
\setcounter{figure}{11}
\begin{figure}
\hspace{2cm}
\centerline{\psfig{figure=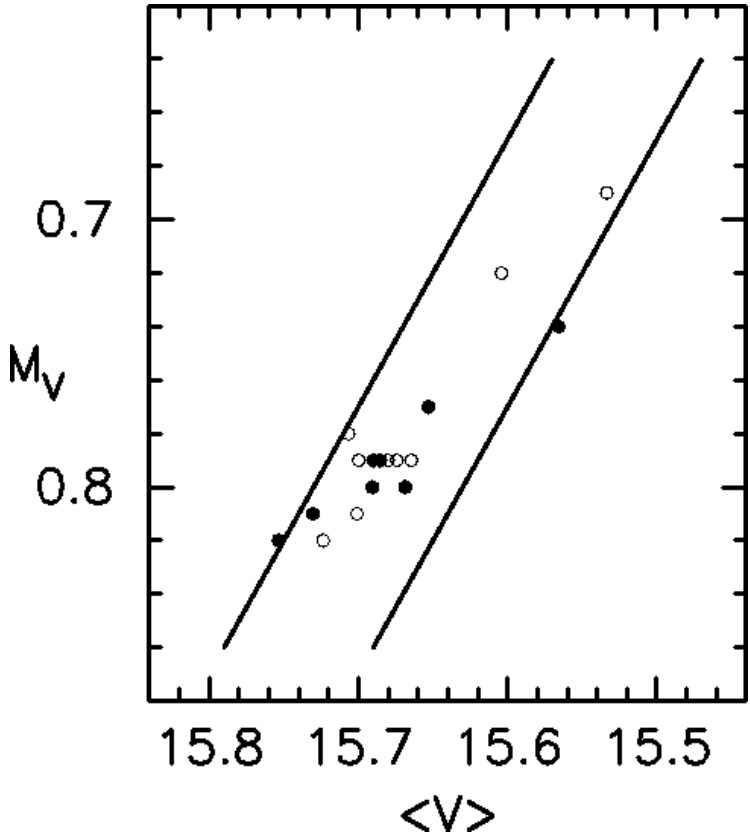,height=11cm}} 
\caption{
The absolute magnitude derived from the KJ96 equation relating
$M_V$ to period and the Fourier parameters $\phi_{31}$ and $A_1$ plotted
against mean $V$  for the RRab stars in M3. The open circles represent the
stars in the field north of the cluster centre and the solid circles
are for the south field.
The envelope lines have a slope of unity and are separated
by $M_V=0.1$ which represents the uncertainty in the KJ calibration.
}
\end{figure}
\clearpage
\setcounter{figure}{12}
\begin{figure}
\hspace{2cm}
\centerline{\psfig{figure=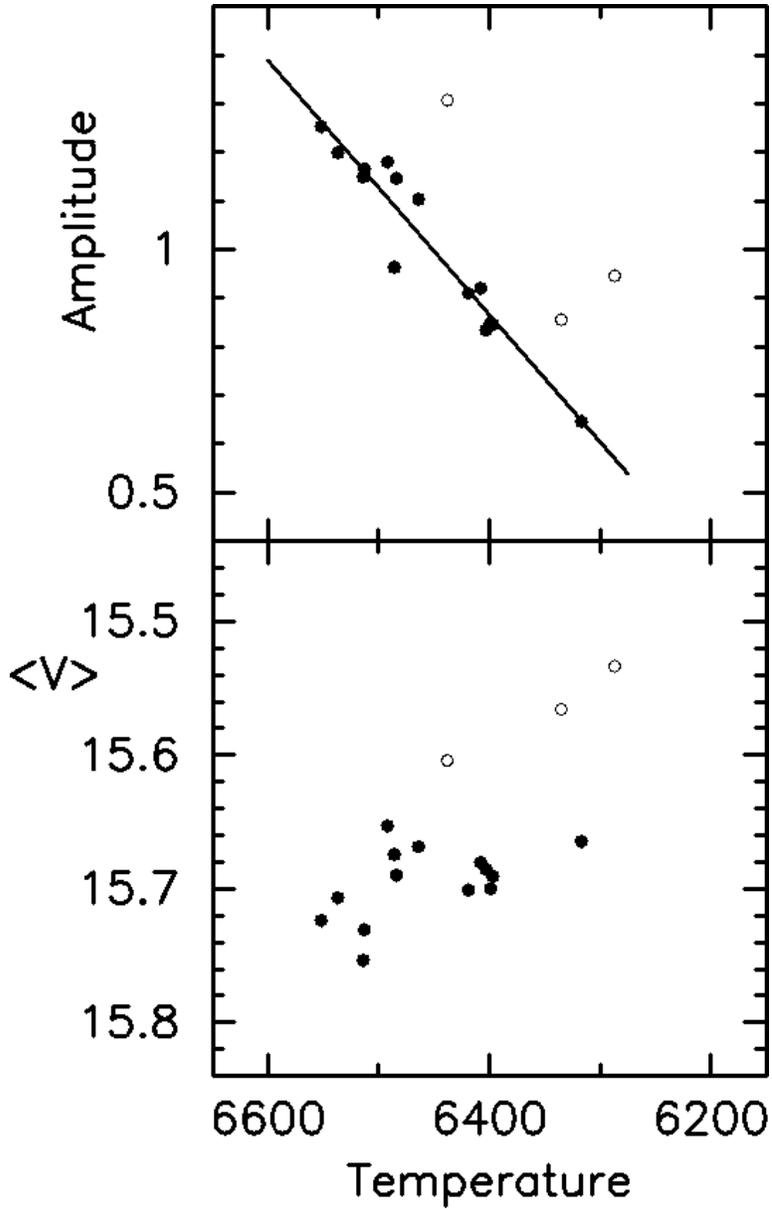,height=16cm}} 
\caption{
Upper panel: The $V$ amplitude versus the temperature listed 
in Table 5
for the RRab stars with `regular' light curves. Open circles are plotted
for the three stars V14, V65 and V104 which are
the three brightest stars in Figure 11 and also the three
stars that have large amplitudes
for their periods compared with the other `regular' RRab stars in Figure 8.
The remaining stars are plotted as solid circles and the straight line is a
least squares fit through the solid circles.   
Lower panel: The mean $V$ magnitude versus temperature for the same stars.
The temperatures of Table 5 were calculated from the pulsation period and the
Fourier parameters  $\phi_{31}$, $\phi_{41}$ and $A_1$. 
}
\end{figure}
\clearpage
\setcounter{figure}{13}
\begin{figure}
\hspace{2cm}
\centerline{\psfig{figure=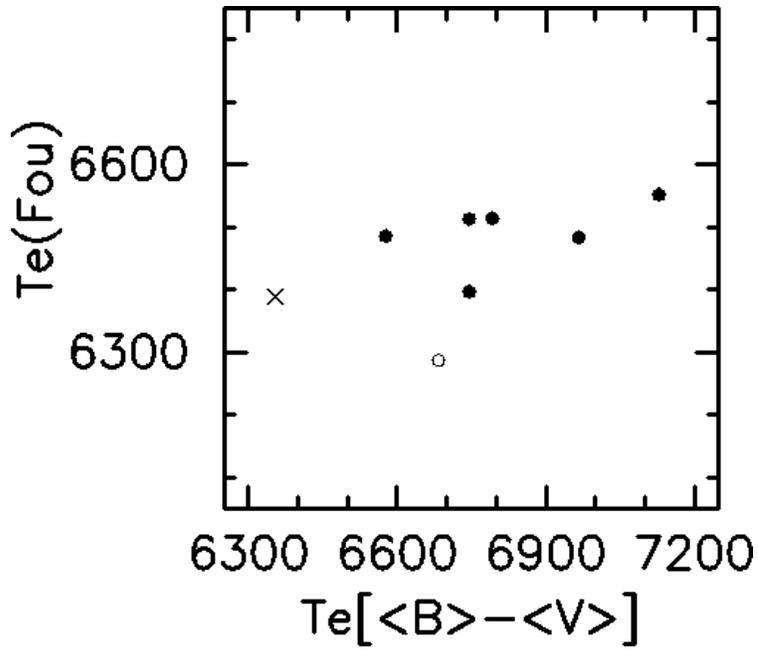,height=8.5cm}} 
\caption{
The temperatures of Table 5 ($T_e$(Fou)) plotted against the
 temperatures derived by Sandage (1981b)  from [$<B>-<V>$]
for the RRab stars V1, V18, V34, V51, V65, V74, V84 and V90. V65 is plotted
as an open circle and V84 as a cross.
}
\end{figure}
\bsp 

\label{lastpage}
\end{document}